# Spin Hall torque driven chiral domain walls in magnetic heterostructures

Jacob Torrejon[1,2] and Masamitsu Hayashi[1]

[1]National Institute for Materials Science, Tsukuba 305-0047, Japan

[2]Unité Mixte de Physique CNRS/Thales, 91767 Palaiseau, France

Email: jacob.torrejon-diaz@thalesgroup.com, hayashi.masamitsu@nims.go.jp

**Abstract**

The motion of magnetic domain walls in ultrathin magnetic heterostructures driven by current via the spin Hall torque is described. We show results from perpendicularly magnetized CoFeB|MgO heterostructures with various heavy metal underlayers. The domain wall moves along or against the current flow depending on the underlayer material. The direction to which the domain wall moves is associated with the chirality of the domain wall spiral formed in these heterostructures. The one-dimensional model is used to describe the experimental results and extract parameters such as the Dzyaloshinskii-Moriya exchange constant which is responsible for the formation of the domain wall spiral. Fascinating effects arising from the control of interfaces in magnetic heterostructures are described.

**Introduction**

Spin transfer torque (STT) [1,2] has enabled control of the magnetization direction in magnetic nanostructures. In magnetic nanowires, STT allows domain walls, boundaries between magnetic domains, to be moved along the wire by current[3-6]. Such current controlled motion of domain walls is the key technology of the "Racetrack memory[7]", a storage class non-volatile memory that may alter the landscape of storage devices. It has been reported that a series of domain walls can indeed be moved in sync with the application of current pulses[8-10]. Many of the underlying physics of STT driven domain wall motion have been uncovered[11-13] and some prototype devices were demonstrated recently[14,15].

The current driven motion of domain walls via STT can be considered as a bulk effect: the effect of layers adjacent to the magnetic layer is negligible in most cases. Recently, however, exciting new phenomena are being discovered in perpendicularly magnetized ultrathin magnetic heterostructures in which the neighboring layers seem to play a significant role in moving domain walls. First, it has been reported that domain walls can be moved along the current flow, opposite to STT driven motion of domain walls, in an ultrathin magnetic layer sandwiched between heavy metal layers or between a heavy metal layer and an insulating oxide layer[10,16,17]. More recently, it has been reported that domain walls move either along or against the current flow depending on the material and stacking order of the heterostructures[18-23]. These results suggest the importance of the interface(s) of the magnetic layer. Theoretically, Thiaville *et al*. have suggested that domain walls, provided that they are Neel-type, can be driven by spin current generated in a neighboring layer that diffuse and impinge upon the magnetic layer [24]. The spin Hall effect (SHE) can generate such spin current, large enough to alter the magnetization direction of a magnetic layer placed to next to the heavy metal layer[25-27]. The torque exerted on the magnetic moments by the impinging spin current is typically called the "spin Hall torque[25]", which is to be distinguished from conventional spin transfer torque since spin orbit coupling plays a role in generating the spin current and possibly the torque.

In a typical perpendicularly magnetized wire, the stable domain wall structure is a Bloch wall[28], in which the magnetization of the domain wall points parallel to the domain boundary. If the width of the wire is sufficiently reduced, magnetostatic energy can force the magnetization of the domain wall to point normal to the boundary (i.e. along the wire's long axis) and form a Neel wall. Such transition in the wall structure has been observed, for example, in perpendicularly magnetized Co|Ni multilayers[12]. The magnetization direction of the Neel walls in narrow width wires is likely to be random. However, to move series of domain walls in sync using the spin Hall torque, the magnetization direction of neighboring domain walls need to alternate, i.e. the chirality of the wall needs to be identical for all domain walls within the wire. This is not possible just by reducing the wire's width. Such chiral domain walls[29] can be realized if the Dzyaloshinskii-Moriya interaction (DMI)[30,31], an anti-symmetric exchange interaction present in bulk crystals with broken inversion symmetry[32,33] or at surfaces[34] and interfaces [19,20,22,23,35-38], is introduced into the system. The coexistence of the spin Hall effect and the DMI enables motion of multiple domain walls in sync along or against the current flow[24,39].

In this article, we describe spin Hall torque driven chiral domain walls in ultrathin magnetic heterostructures. The modified Landau Lifshitz Gilbert (LLG) equation, which includes both spin transfer torque and the spin Hall torque, is used to derive the one-dimensional (1D) model[40] of a domain wall. The 1D model is used to analyze the dependence of domain wall velocity on external magnetic field, which allows extraction of the Dzyaloshinskii-Moriya (DM) exchange constant of the system. Experimental results on current driven domain wall motion in magnetic heterostructures are shown to compare them with the 1D model.

1. **The one-dimensional model of domain walls**

The Landau Lifshitz Gilbert (LLG) equation that includes the adiabatic/non-adiabatic STT and the spin Hall torque can be written as the following[5,41].

$$\frac{\partial \hat{m}}{\partial t} = -\gamma \hat{m} \times \left( -\frac{\partial E}{\partial \vec{M}} + a_J (\hat{m} \times \hat{p}) + b_J \hat{p} \right) + \alpha \hat{m} \times \frac{\partial \hat{m}}{\partial t} - u \left( \hat{j} \cdot \vec{\nabla} \right) \hat{m} + \beta u \hat{m} \times \left( \hat{j} \cdot \vec{\nabla} \right) \hat{m}, \quad (1)$$

where $\hat{m}$ and $\hat{j}$ represent unit vectors of the magnetic layer's spatially varying magnetization and the current flow along the wire, respectively. $t$ is time. $u = -\frac{\mu_B P}{e M_S} J$ represents the adiabatic STT term, where $\mu_B$ and $e$ are the Bohr magnetron and the electron charge (we define $e>0$ for convenience), $P$ and $M_S$ are the current spin polarization and saturation magnetization of the magnetic layer and $J$ is the current density that flows through the magnetic layer. $\beta$ is the non-adiabatic STT term. $\gamma$ is the gyromagnetic ratio and $\alpha$ is the Gilbert damping constant. $E$ is the total magnetic energy of the system that includes, for example, the demagnetization, anisotropy, exchange, magneto-elastic and the Zeeman energies. $\vec{M} \equiv M_S \hat{m}$.

In Eq. (1), the spin Hall torque is included in the form of effective magnetic field: $a_J$ and $b_J$ correspond to the damping-like (Slonczweski-Berger[1,2]) and the field-like[41] components of the effective field, respectively. $\hat{p}$ is an unit vector representing the spin direction of the electrons impinging upon the magnetic layer via the spin Hall effect.

We use the one-dimensional (1D) model of a domain wall derived by Malozemoff and Slonczweski[40]. Definition of the coordinate system is shown in Fig. 1. The following ansatz is used for the magnetization profile along the wire:

$$\theta(x,t) = 2\arctan\left[\exp\left(\frac{x-q(t)}{\Delta}\right)\right], \quad \varphi(x,t) = \psi(t): \uparrow\downarrow \text{ wall } (\Gamma=1), \quad (2a)$$

$$\theta(x,t) = \pi - 2\arctan\left[\exp\left(\frac{x-q(t)}{\Delta}\right)\right], \quad \varphi(x,t) = -\psi(t): \downarrow\uparrow \text{ wall } (\Gamma=-1). \quad (2b)$$

where $\theta$ and $\varphi$ represent the polar and azimuthal angle of the magnetization $\hat{m}$ that depends on position $x$ and time $t$. $q$ and $\psi$ are the center position and the magnetization angle of the wall, respectively. $\psi=0, \pi$ and $\psi=\pi/2, 3\pi/2$ correspond to perfect Neel and Bloch walls, respectively. $\Delta$ is the domain wall width parameter: the physical width of the wall is $\sim\pi\Delta$. We assume $\psi$ is constant within the domain wall and $\Delta$ is both time and space independent. After some algebra, two coupled ordinary differential equations are obtained from Eq. (1):

$$(1+\alpha^2)\frac{\partial q}{\partial t} = \Delta\left[-\frac{1}{2}\gamma H_K \sin 2\psi - \frac{\pi}{2}\gamma\Gamma(H_Y + b_J)\cos\psi + \frac{\pi}{2}\gamma(H_X + H_{DM})\sin\psi + \frac{u}{\Delta}\right]$$
$$+ \alpha\Delta\left[-\frac{\gamma}{2M_S}\left(\frac{\partial\sigma_{PIN}}{\partial q}\right) + \gamma\Gamma H_Z + \frac{\pi}{2}\gamma\Gamma a_J\cos\psi + \beta\frac{u}{\Delta}\right], \quad (3a)$$

$$(1+\alpha^2)\frac{\partial\psi}{\partial t} = -\alpha\left[-\frac{1}{2}\gamma H_K \sin 2\psi - \frac{\pi}{2}\gamma\Gamma(H_Y + b_J)\cos\psi + \frac{\pi}{2}\gamma(H_X + H_{DM})\sin\psi + \frac{u}{\Delta}\right]$$
$$+ \left[-\frac{\gamma}{2M_S}\left(\frac{\partial\sigma_{PIN}}{\partial q}\right) + \gamma\Gamma H_Z + \frac{\pi}{2}\gamma\Gamma a_J\cos\psi + \beta\frac{u}{\Delta}\right]. \quad (3b)$$

$H_K$ is the magnetic anisotropy field associated with the domain wall magnetization: it corresponds to the field needed to cause transition between Neel and Bloch walls. The effect of a pinning potential is described by the term with $\partial\sigma_{PIN}/\partial q$: $\sigma_{PIN}$ denotes the wall pinning potential energy density. $H_Z$, $H_X$ and $H_Y$ correspond to the out of plane, in-plane longitudinal (along the current flow direction) and in-plane transverse (transverse to the current flow direction) external fields, respectively. $\Gamma$ represents the domain pattern; $\Gamma=+1$ for ↑↓ wall and $\Gamma=-1$ for ↓↑ wall. The effect of DMI on the wall magnetization is included[19,20,22-24] as a longitudinal offset field $H_{DM}$ directed along the wire's long axis. $H_{DM} = \dfrac{\Gamma D}{M_S \Delta}$, where $D$ is the DM exchange constant.

When the spin Hall torque is the dominant driving force in moving domain walls, the steady state magnetization tilt angle during the current application approaches $\psi \sim \pm \pi/2$. In the absence of pinning ($\sigma_{PIN}=0$), one can then linearize Eqs. (3a) and (3b) around $\psi \sim \pm \pi/2$ to obtain the following analytical form of the domain wall velocity[19,20,23]:

$$v_{DW} = \frac{\frac{\pi}{2}\Gamma a_J}{\pm \frac{\pi}{2}\Gamma a_J + \alpha\left[H_K \pm \frac{\pi}{2}\Gamma(H_Y+b_J)\right]}\left[\frac{\pi}{2}\gamma\Delta(H_X+H_{DM})\pm u\right]$$
$$+ \frac{H_K \pm \frac{\pi}{2}\Gamma(H_Y+b_J)}{\pm \frac{\pi}{2}\Gamma a_J + \alpha\left[H_K \pm \frac{\pi}{2}\Gamma(H_Y+b_J)\right]}(\gamma\Delta H_Z + \beta u) \quad (4)$$

The ±sign corresponds to the case when $\psi$ approaches $\pm\pi/2$. From hereon, we assume $H_Z=0$ and $\beta=0$ since the out of plane field $H_Z$ is set near zero during the measurements of current driven domain wall velocity and it has been reported recently that contribution form the non-adiabatic STT term is negligible in perpendicular magnetized CoFeB ultrathin films[42]. For spin Hall torque driven domain wall motion, the ±sign in Eq. (4) is given by $\text{sgn}(\Gamma \cdot a_J)$. Replacing the ±sign with $\text{sgn}(\Gamma \cdot a_J)$ and substituting $H_Z=\beta=0$ in Eq. (4) give:

$$v_{DW} = \frac{\frac{\pi}{2}\gamma\Delta\Gamma a_J}{|a_J| + \alpha\left[\frac{2}{\pi}H_K + \text{sgn}(a_J)(H_Y + b_J)\right]}\left[H_X + \left(H_{DM} + \frac{2}{\pi}\text{sgn}(\Gamma a_J)\frac{u}{\gamma\Delta}\right)\right] \quad (5)$$

Equation (5) shows that wall velocity is linear against the in-plane longitudinal field ($H_X$). The intercept between this linear line and the horizontal axis with $v_{DW}=0$ gives the compensation field $H_X^*$ which contains information of $H_{DM}$, i.e.

$$H_X^* = -\left(H_{DM} + \frac{2}{\pi}\text{sgn}(\Gamma a_J)\frac{u}{\gamma\Delta}\right) \quad (6)$$

If STT contribution to the wall motion is absent (i.e. $u=0$), then $H_X^*$ is exactly equal to $-H_{DM}$. However, in the heterostructures studied here (X|CoFeB|MgO), it turns out that we cannot neglect the influence of $u$ in calculating $H_{DM}$ from $H_X^*$ since a fraction of current flows into the magnetic layer.

The dependence of the wall velocity against the in-plane transverse field ($H_Y$) can be extracted from Eq. (4) if $H_Y$ is small compared to the domain wall anisotropy field $H_K$. Assuming $H_Y \ll H_K$ and substituting $H_X=0$ in Eq. (4) give the following relation between the velocity and $H_Y$:

$$v_{DW} \sim \Gamma H_X^* \frac{\left(\frac{\pi}{2}\right)^2 \gamma\Delta\alpha|a_J|}{\left[|a_J| + \alpha\left(\frac{2}{\pi}H_K + \text{sgn}(a_J)b_J\right)\right]^2}\left[H_Y - \frac{|a_J| + \alpha\left(\frac{2}{\pi}H_K + \text{sgn}(a_J)b_J\right)}{\alpha\,\text{sgn}(a_J)}\right] \quad (7)$$

Equation (7) shows that in the small $H_Y$ limit where the wall velocity is linear with $H_Y$, the slope of $v_{DM}$ versus $H_Y$ contains information of $H_{DM}$ via $H_X^*$. The magnitude of the slope includes contribution from other parameters ($a_J$, $H_K$, $\alpha$ and $b_J$ if any), thus hindering direct evaluation of $H_{DM}$. The sign of the slope, determined by the product of the wall type and the compensation field, i.e. $\Gamma H_X^*$, represents the domain wall chirality if contribution from the STT term is absent (i.e. $u=0$): $\Gamma H_X^* > 0$ for left-handed and $\Gamma H_X^* < 0$ for right-handed chirality, respectively.

In the presence of pinning, the threshold current to move (i.e. depin) a domain wall is obtained by assuming a parabolic pinning potential[43]

$$\sigma_{PIN} = \frac{V}{q_0} q^2 \vartheta(|q| - q_0) \qquad (8)$$

where $V$ and $q_0$ are the depth and width of the potential well and $\vartheta(q)$ is a Heaviside step function. The threshold current is defined as the current needed to shift the equilibrium position of a domain wall out from the well. As described below, it turns out that the threshold current depends on whether the domain wall is a Neel wall or a Bloch-like wall[23]. In order to stabilize a Neel wall, the longitudinal offset field $H_{DM}$ needs to overcome the anisotropy field associated with the wall, i.e.

$$H_{DM}^{NEEL} = \frac{2}{\pi} H_K \qquad (9)$$

If $H_{DM}$ is smaller than $H_{DM}^{NEEL}$, the wall becomes Bloch-like since the wire width is large enough to stabilize Bloch walls. Reducing the wire width to sub-hundred nanometers will, in general, allow stabilization of Neel walls via gain in magnetic shape anisotropy ($H_K$ will then be negative in Eq. (2)). As noted above, one can introduce Neel walls by scaling down the wire width but the chirality of the domain walls will be more or less random, which does not allow in sync motion of multiple domain walls with the spin Hall torque.

The threshold spin Hall effective field for depinning the Bloch and Neel-like walls are expressed as[23]:

$$a_J^C = \frac{4}{\pi^2} H_P \frac{H_K}{H_{DM}} \qquad (|H_{DM}| \leq H_{DM}^{NEEL}) \qquad (10a)$$

$$a_J^C = \frac{2}{\pi} H_P \qquad \left( |H_{DM}| \geq H_{DM}^{NEEL} \right) \qquad (10b)$$

where $H_P$ is the propagation (depinning) field which depends on the strength of pinning: i.e. $H_P = V/M_S$. $a_J^C$ decreases as the domain wall changes its magnetization direction from a Bloch-like configuration ( $|H_{DM}| \leq H_{DM}^{NEEL}$ ) to that of a Neel wall. For a perfect Neel wall ($|H_{DM}| \geq H_{DM}^{NEEL}$), $a_J^C$ depends only on $H_P$.

## 2. Experimental results

### A. Sample preparation and experimental methods

We study magnetic heterostructures consisting of Substrate|$d$ X|1 Co$_{20}$Fe$_{60}$B$_{20}$|2 MgO|1 Ta (units in nm) using different heavy metal underlayers (X: Hf, Ta, TaN and W). Films are deposited on thermally oxidized Si(001) substrates (SiO$_2$: 100 nm thick) using magnetron sputtering. The TaN underlayer is formed by reactively sputtering Ta in Ar gas atmosphere mixed with small amount of N$_2$ gas. The atomic composition of TaN is determined by Rutherford backscattering spectroscopy and contains 52±5 at% of N. Films are annealed at 300 °C for 1 h in vacuum. The thickness of the underlayer X is linearly varied across the substrate to study the thickness dependence of various parameters. The film thickness of underlayer X is calibrated by comparing the resistance (and the anomalous Hall resistance) of a Hall bar patterned on the wedge film with that of a flat film with constant thickness of X. The magnetic properties of the heterostructures are studied in vibrating sample magnetometry (VSM) using the flat films.

Hall bars and wires are patterned from the films using optical lithography and Ar ion etching. Subsequent lift-off process is used to form the electrical contact 10 Ta|100 Au (units in

nanometers). The Hall bars used to study current induced (spin Hall) effective fields are 10 μm wide and 30-60 μm long whereas the wires for domain wall motion experiments are 5μm wide and 30 μm long. Details of the device preparation, magnetic, transport and structural properties of the heterostructures can be found in Refs [23,44,45].

## B. Evaluation of the current induced effective field

Adiabatic harmonic Hall measurements[44,46,47] are used to estimate the current induced effective field arising from the spin Hall torque. A low frequency (~500 Hz) sinusoidal voltage (amplitude: $V_{IN}$) is applied to the Hall bar and the in-phase first ($V_\omega$) and the out of phase second ($V_{2\omega}$) harmonic Hall voltages are measured using lock-in amplifiers[44]. An in-plane field is applied along ($H_X$) or transverse to ($H_Y$) the current flow. The field dependence of the harmonic Hall voltages provides information of the longitudinal and transverse effective fields directed along the x and y axes, respectively[48]:

$$\Delta H_X = -2\frac{(B_X \pm 2\xi B_Y)}{1-4\xi^2}, \tag{11a}$$

$$\Delta H_Y = -2\frac{(B_Y \pm 2\xi B_X)}{1-4\xi^2}. \tag{11b}$$

We define $B_X \equiv \frac{\partial V_{2\omega}}{\partial H_X}\bigg/\frac{\partial^2 V_\omega}{\partial H_X^2}$, $B_Y \equiv \frac{\partial V_{2\omega}}{\partial H_Y}\bigg/\frac{\partial^2 V_\omega}{\partial H_Y^2}$ and $\xi \equiv \frac{\Delta R_P}{\Delta R_A}$, where $\Delta R_P$ and $\Delta R_A$ are the planar Hall and the anomalous Hall contributions to the Hall resistance, respectively.

Both components of the effective field ($\Delta H_X$ and $\Delta H_Y$) scale linearly with the amplitude of the sinusoidal voltage ($V_{IN}$) at low excitation. $V_{IN}$ can be converted to the current density that flows through the underlayer ($J_N$) using the resistance of the wire, the resistivity and the thickness of the underlayer X and the CoFeB layer. We fit $\Delta H_{X(Y)}$ vs. $J_N$ with a linear function to obtain the effective field per unit current density, $\Delta H_{X(Y)}/J_N$.

Results of $\Delta H_{X(Y)}/J_N$ measured for the TaN underlayer films (Sub.|d TaN|1 CoFeB|2 MgO|1 Ta) are shown in Fig. 1(b) and 1(c) as a function of the TaN underlayer thickness. Solid and open symbols correspond to $\Delta H_{X(Y)}/J_N$ when the equilibrium magnetization direction is pointing along +z and −z, respectively. As evident, both components of the effective field increase with increasing underlayer thickness. This is consistent with the picture of spin Hall torque, in which the magnitude of the torque scales with the underlayer thickness up to its spin diffusion length[25,49-51], above which the torque saturates since spin current generated far from the underlayer|magnetic layer interface (distance larger than the spin diffusion length) loose its spin information before reaching the interface.

Figure 1(b) and 1(c) show that $\Delta H_Y/J_N$ is the same regardless of the magnetization direction whereas $\Delta H_X/J_N$ changes its direction when the magnetization direction is reversed. These results illustrate the correspondence between $\Delta H_X$ ($\Delta H_Y$) and $a_J$ ($b_J$) in Eqs. (1), that is:

$$\begin{aligned} \Delta H_X &\sim a_J (\hat{m} \times \hat{p}), \\ \Delta H_Y &\sim b_J \hat{p}. \end{aligned} \quad (12)$$

As described after Eq. (1), $\hat{p}$ is the spin direction of the electrons entering the ferromagnetic (FM) layer via the spin Hall effect that takes place in the non-magnetic (NM) underlayer (see Fig. 1(a)). $\Delta H_X$ and $\Delta H_Y$ can thus be considered as the damping-like and field-like components, respectively. Recently it has been reported that there are components of the current induced effective field that are dependent on the angle between the magnetization and current[47,52,53]. Although these components are reported to be non-negligible in many heterostructures, here we do not consider these angle dependent terms in describing domain wall dynamics and thus assume Eq. (12) to hold.

The relative size of the damping-like and field-like terms depends on the materials and thickness of the underlayer and the magnetic layer. For Ta and TaN underlayer films, we find that the field-like term ($\Delta H_Y/J_N$) is 2-3 times larger than the damping-like term ($\Delta H_X/J_N$)[23,44]. The direction of the damping like term is consistent with that originally predicted by Slonczewski if one assumes the spins that impinge upon the magnetic layer point along the direction dictated by the sign of the spin Hall angle[25,54]. The direction of the field-like term is opposite to the incoming spin direction, which is rather counter-intuitive. The size and the direction of the field-like term is said to depend on how the electrons reflect and/or transmit the NM|FM interface[55-57].

According to Eq. (5), the domain wall velocity is proportional to the size of the damping like term $a_J$. Thus, from Eq. (12), it is $\Delta H_X/J_N$ that provides the driving force in moving domain walls.

**C. Measurements of domain wall velocity**

An optical microscope image of a typical wire used to evaluate domain wall velocity is shown in Fig. 2(a). The wire width is ~5 μm and we study propagation of domain wall(s) over a distance of ~30 μm using Kerr microscopy imaging. A pulse generator, which can apply constant amplitude voltage pulse, is connected to the wire. Definition of the coordinate axes is the same with that shown in Fig. 1(a). Positive current corresponds to current flow along the +x direction. Exemplary hysteresis loops of the wire are shown in Fig. 2(b) for two films with different underlayer material (Hf and W). Hysteresis loops are obtained by the Kerr microscopy images of the wire captured during an out of plane magnetic field sweep ($H_Z$). For all films analyzed here, the switching field is governed by the field needed to nucleate reverse domains and ranges between ~100 to ~500 Oe.

Preparation of a domain wall (or domain walls) in the wire is carried out by applying large amplitude voltage pulses. First, the CoFeB layer is uniformly magnetized by applying large enough out of plane field ($H_Z$). The field is then removed and a large amplitude voltage pulse is applied to the wire, which can trigger magnetization reversal[58,59]. This typically results in nucleation of one or two domain walls within the wire. In some cases, additional field is applied after the voltage pulse application in order to form appropriate domain structure. The pulse amplitude needed to trigger magnetization reversal is in general, larger than that needed to move domain walls: thus this sets the upper limit of the applicable pulse amplitude for studying current driven domain wall motion.

For studying current induced domain wall motion, we use ~100 ns long voltage pulses (shorter pulses, 20-50 ns long, are used when the velocity of the wall becomes large). Kerr images are captured before and after the application of the voltage pulse(s) to estimate the distance the wall traveled. Figures 2(c) and 2(d) show such Kerr images for the two films shown in Fig. 2(b). The upper image shows the initial state with two domain walls in the wire; the lower image illustrates the magnetic state after the application of the voltage pulse. As evident, the two domain walls have moved in sync with the application of the voltage pulse. The direction to which the walls move depends on the film structure, as will be discussed below.

Figure 2(e) and 2(f) show successive positions of the two domain walls shown in Figs. 2(c) and 2(d), respectively, as voltage pulses are applied to the wire. The cumulated pulse length is shown in the horizontal axis to extract domain wall velocity from this plot. When the driving force is large enough, domain walls can be driven along the wire without pinning. However, in some circumstances when the pinning is strong or when the driving force is weak, domain walls can get locally pinned. An example of such pinning is shown in Fig. 2(e) where the position of one of the domain walls does not change with the application of voltage pulses (the corresponding cumulated pulse length is ~1-2 μs). In this case, after application of a few voltage pulses, the wall depins and restarts its motion along the wire. The average domain wall velocity is estimated by fitting the wall position as a function of cumulated pulse length with a linear function. We exclude cases when the domain wall is locally pinned. The average value of the slope of the solid lines in Figs. 2(e) and 2(f) gives the domain wall velocity.

Figure 3 shows the dependence of domain wall velocity on the current density $J_N$ flowing through the NM underlayer for four different film structures. Positive velocity means the domain wall moves along the +x direction. The threshold current needed to trigger the wall motion depends on the film structure (we do not consider creep motion here). According to Eq. (10), domain walls can be moved if the spin Hall effective field $a_J$ exceeds $a_J^C$, which depends on the wall structure (Neel or Bloch-like) and the domain wall propagation field $H_P$. $H_P$ is studied using Kerr images and is defined as the average minimum out of plane field needed to move a domain wall along the wire. $H_P$ is listed together in each panel in Fig. 3. The threshold effective field $a_J^C$ can be calculated using the threshold current density $J_N^C$ obtained from Fig. 3 and the effective field per unit $J_N$ from Fig. 1. For example, for 6.6 nm thick TaN underlayer film, $J_N^C \sim 0.3 \times 10^8$ A/cm$^2$ from Fig. 3(c) and $|\Delta H_X/J_N| \sim 235$ Oe/($1 \times 10^8$ A/cm$^2$) from Fig. 1(c); thus $a_J^C = J_N^C \times (\Delta H_X / J_N)$ ~70 Oe (we assume Eq. (12) holds here). This is larger than the propagation field ($H_P$~30 Oe), indicating that the domain wall is not a perfect Neel wall. Using Eq. (10a), these values give $H_{DM}/(\pi H_K/2) \sim 0.27$ that corresponds to the direction cosine of the magnetization angle with respect to the wire's long axis (here, along the x-axis).

**D. Determination of the magnetic chirality**

Figure 3 shows that the domain wall moves along with the current flow for the TaN and W underlayer films whereas it moves against it for the Hf and Ta underlayer films. The direction to which a domain wall moves is determined by the sign of the spin Hall angle and the wall chirality[19,20,22-24], i.e. the DM exchange constant. The effective field measurements (e.g. Fig. 1) can, in general, provide the sign of the spin Hall angle. We find that the sign of the spin Hall angle is the same for the underlayer materials used here (Hf, Ta, TaN and W)[23]. Note that estimation of the size of the spin Hall angle from such measurements is difficult as the size of the effective field depends on the product of the spin Hall angle and the spin mixing conductance of the NM|FM interface[57]. As shown below, here it is the DM exchange constant that differs depending on the underlayer material and consequently changes the direction to which a domain wall moves.

The in-plane field dependence of the domain wall velocity is shown in Fig. 4 for the TaN underlayer film. The velocity scales linearly with the in-plane field in this field range. For the longitudinal field ($H_X$) sweep the slope of this linear relationship changes its sign when the current direction is reversed or if the wall type is changed between ↑↓ and ↓↑ walls. In contrast, the slope is the same for ↑↓ and ↓↑ walls as well as for positive and negative currents for the transverse field ($H_Y$) sweep. This trend agrees with the 1D model: according to Eq. (5) and (7), the sign of the slope of $v_{DW}$ vs. $H_X$ and $v_{DW}$ vs. $H_Y$ is given by $\Gamma a_J$ and $\Gamma H_X^*$ ($\Gamma$ is 1 for ↑↓ wall and -1 for ↓↑ walls), respectively. As $a_J$ depends on the current direction, the sign of the slope for $v_{DW}$ vs. $H_X$ depends on the wall type and the current flow direction. The slope for $v_{DW}$ vs. $H_Y$ does not change its sign with the wall type since $H_X^*$ also depends on $\Gamma$ via $H_{DM}$ (see Eqs. (6) and (7)). The negative slope of $v_{DW}$ vs. $H_Y$ in Figs 4(a) and 4(b) thus indicates that the domain walls are right handed for this film (TaN underlayer). The longitudinal compensation field $H_X^*$ can be found from the plots shown in Fig. 4(c) and 4(d).

Figure 5 shows the longitudinal field dependence of the wall velocity for three different underlayer films. The sign of the compensation field $H_X^*$ is opposite for Hf, Ta and TaN, W underlayer films (see Fig. 4c, 4d for $H_X^*$ of the TaN underlayer films), which is in agreement with the direction to which domain walls move with current, as shown in Fig. 3 (the thickness of the underlayer is slightly different for Fig. 3 and Figs. 4, 5). Equation (6) shows that $H_X^*$ is constant regardless of the pulse amplitude (or $J_N$) if contribution from the STT term is small ($u$~0). This applies for the Hf and W underlayer films. However $H_X^*$ changes appreciably when $J_N$ is varied for the Ta underlayer films.

To extract the DM exchange constant properly using Eq. (6), the adiabatic spin torque term $u$ and the domain wall width parameter $\Delta$ needs to be determined. First, the current density ($J$) that flows into the CoFeB layer needs to be substituted in the expression of $u$. $J$ is calculated using the thickness and resistivity of each layer: $\rho_{CoFeB}$ = 160, $\rho_{Hf}$ = 199, $\rho_{Ta}$ = 189, $\rho_{TaN}$ = 375 and $\rho_W$ = 124 $\mu\Omega \cdot cm$[23]. The saturation magnetization depends on the material and thickness of the underlayer[45]: here for simplicity we assume $M_S$~1500 emu/cm$^3$, which is close to that of bulk $Co_{20}Fe_{60}B_{20}$. The current spin polarization of the CoFeB layer has been reported to be ~0.7 in a similar system[42]: we use this value as a median and use the error bars to show the range of the DM exchange constant when $P$ is varied from 0 to 1. The domain wall width parameter is inversely proportional to the effective magnetic anisotropy energy $K_{EFF}$ of the film, i.e. $\Delta = \sqrt{A/K_{EFF}}$, where $A$ is the exchange stiffness constant. $K_{EFF}$ is determined from the magnetization hysteresis loops[23] and we use $A$~3.1 erg/cm$^3$ estimated from a different study reported previously[60].

In Fig. 6, we show the compensation field $H_X^*$ as a function of the underlayer thickness for four film structures. The background color indicates the direction to which a domain wall moves when current is applied. We fit the underlayer thickness dependence of $H_X^*$ using Eq. (6): the change in $K_{EFF}$ with the underlayer thickness is taken into account for the fitting. This fitting assumes that the DM exchange constant does not depend on the thickness of the underlayer, which may not be the case since the thickness can influence the state of interface both structurally and electronically. The fitted values of the DM exchange constant are summarized in Fig. 6(e) for all film structures studied. As evident, the DM exchange constant $D$ changes as the underlayer material is varied. $D$ is negative for Hf underlayer and is nearly zero for Ta. It increases when nitrogen is added to Ta to form TaN, and $D$ takes the largest value here for W underlayer. These results show that the DM exchange constant can be controlled by the NM|FM interface[23].

### 3. Concluding remarks

We have described the underlying physics of current driven domain wall motion in ultrathin magnetic heterostructures. With the introduction of the spin Hall torque and chiral magnetic structure, domain walls can be moved along the wire either along or against the current flow depending on the material and stacking order of the magnetic heterostructure. In order to fully utilize spin Hall torque and chiral magnetic structure to move domain walls formed in ultrathin magnetic heterostructures, it is essential to find a film structure in which the spin Hall torque and the Dzyaloshinskii-Moriya interaction are greatly enhanced. According to the one-dimensional model, Neel walls can be moved with current only when the spin Hall effective field exceeds the wall pinning field. Thus to simultaneously achieve thermally stable domain walls and low threshold current, one needs to find a system in which the spin Hall torque becomes sufficiently large to overcome the large pinning field needed for high thermal stability. This is in contrast to adiabatic spin transfer torque (STT) driven domain walls, where the threshold current is not related to the pinning field. With the engineering of the film stack and materials innovation, however, we hope that this field will further grow and develop viable technologies in the near future.

**Acknowledgements**

We thank J. Sinha for sample preparation and film characterization, J. Kim for the measurements of the spin Hall effective field, and M. Yamanouchi, S. Takahashi, S. Mitani, S. Maekawa, H. Ohno for helpful discussions. This work was partly supported by the Grant-in-Aid (25706017) from MEXT and the FIRST program from JSPS.

**Figure captions**

**Fig. 1** (a) Schematic image of the system. The motion of electrons when the spin Hall effect takes place in the underlayer (e.g. Ta). The current induced effective field arising from the spin Hall torque is shown by the large blue and red arrows, representing their size and direction. (b,c) The *y* (b) and *x* (c) components of the current induced effective field plotted against the TaN underlayer thickness (source: Ref. [23]). Solid and open symbols correspond to magnetization directed along +z and –z, respectively. The effective field is normalized by the current density $J_N$ that flows through the underlayer (TaN here).

**Fig. 2** (a) Optical microscopy image of the wire used to study domain wall velocity. The coordinate axes are shown together. (b) Hysteresis loops obtained by Kerr microscopy imaging. The region of interest of the captured image is converted to numbers to quantify the Kerr intensity. The intensity is plotted as a function of out of plane field ($H_Z$) for two different film structures: W and Hf underlayer films. (c,d) Kerr images of two domain walls before (upper image) and after (lower image) the application of current pulses. The underlayer is Hf and W for the (c) and (d), respectively. (e,f) Successive position of the domain walls, shown in (c) and (d), upon application of voltage pulses plotted as a function of the cumulated pulse length. The symbols represent the positon of the domain wall after application of the following voltage pulse train: (e) 14 V, 50 ns long pulses applied 5 times and (f) 16 V, 50 ns long pulses applied 2 times, each pulse separated by ~10 ms.

**Fig. 3** (a-d) Domain wall velocity plotted as a function of the current density that flows through the underlayer ($J_N$). The thickness and material of the underlayer is shown in each panel. The domain wall propagation field ($H_P$) for the corresponding device is displayed next to the panel. The vertical dotted lines illustrate the difference in the threshold current for different underlayer films. (Source: Ref. [23])

**Fig. 4** (a-d) In-plane field dependence of the domain wall velocity for the TaN underlayer film. The in-plane field is directed along the *y*-axis (a,b) and the *x*-axis (c,d). Solid and open symbols represent results for positive and negative currents. (a,c) show results for ↓↑ walls and (b,d) are for the ↑↓ walls. The current density that flows through the underlayer ($J_N$) is indicated at the top right (pulse amplitude used here is 28 V). The solid and dashed lines are linear fit to the data to extract $H_X^*$.

**Fig. 5** (a-f) Domain wall velocity as a function of in-plane field directed along the *x*-axis ($H_X$) is shown for three different films. (a,b) Hf, (c,d) Ta and (e,f) W underlayer films. Solid and open symbols represent the results for positive and negative currents. Results from two different values of current density flowing through the underlayer are shown using different symbols. The corresponding pulse amplitude is (a,b) 12 and 16 V, (c,d) 35 and 40 V and (e,f) 10 and 14 V. (a,c,e) show results for ↓↑ walls and (b,d,f) are for the ↑↓ walls. The solid and dashed lines are linear fit to the data to extract $H_{X^*}$.

**Fig. 6** (a-d) The compensation field $H_X^*$, i.e. the longitudinal field ($H_X$) at which the velocity becomes zero, plotted as a function of underlayer thickness for (a) Hf, (b) Ta, (c) TaN and (d) W underlayer films (source: Ref. [23]). Solid and open symbols represent ↑↓ and ↓↑ domain walls, respectively. $H_X^*$ is evaluated when the wall is driven either by positive or negative voltage pulses: here, both results are shown together. The background color of each panel indicates the direction to which a corresponding domain wall moves; red: along the current flow, blue: against the current flow. Solid and dashed lines represent fitting using Eq. (6) to estimate the Dzyaloshinskii-Moriya exchange constant $D$. (e) $D$ as a function of underlayer material. The center panel shows $D$ against the atomic concentration of N in TaN. The error bars show the range of $D$ when contribution from spin transfer torque is changed: lower (higher) bound of the error bars corresponds to $P=0$ ($P=1$) and the symbols assume $P=0.7$.

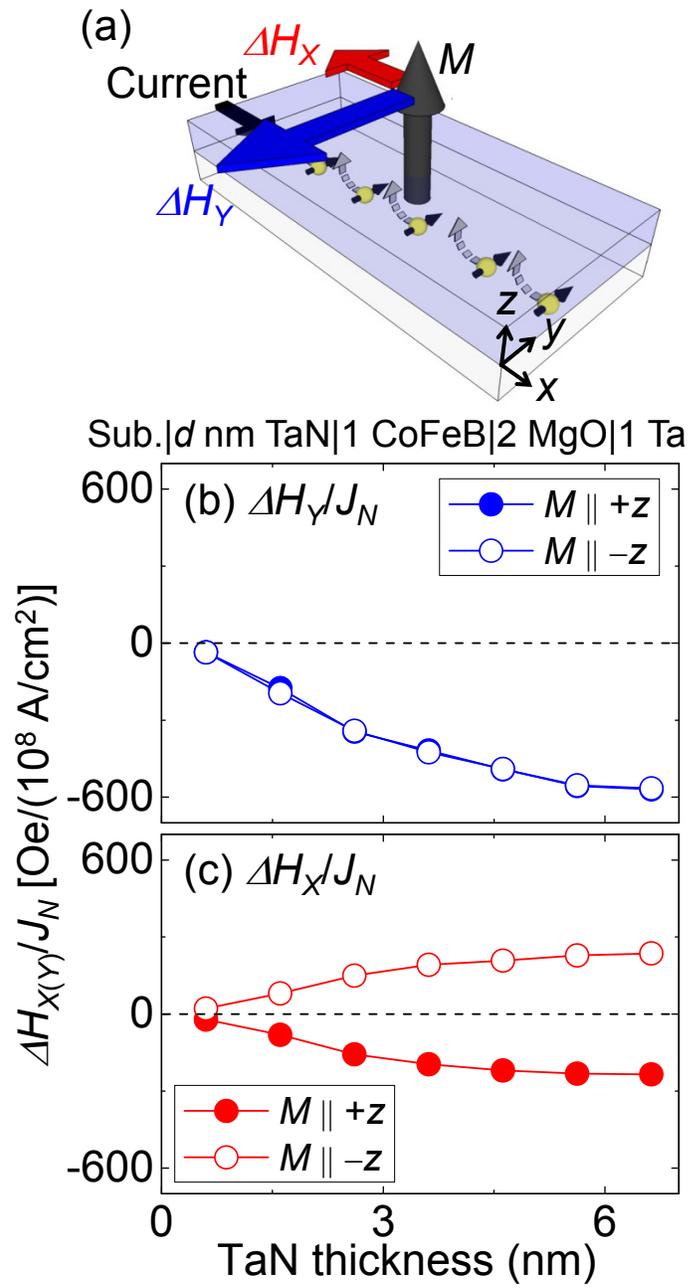

Fig. 1

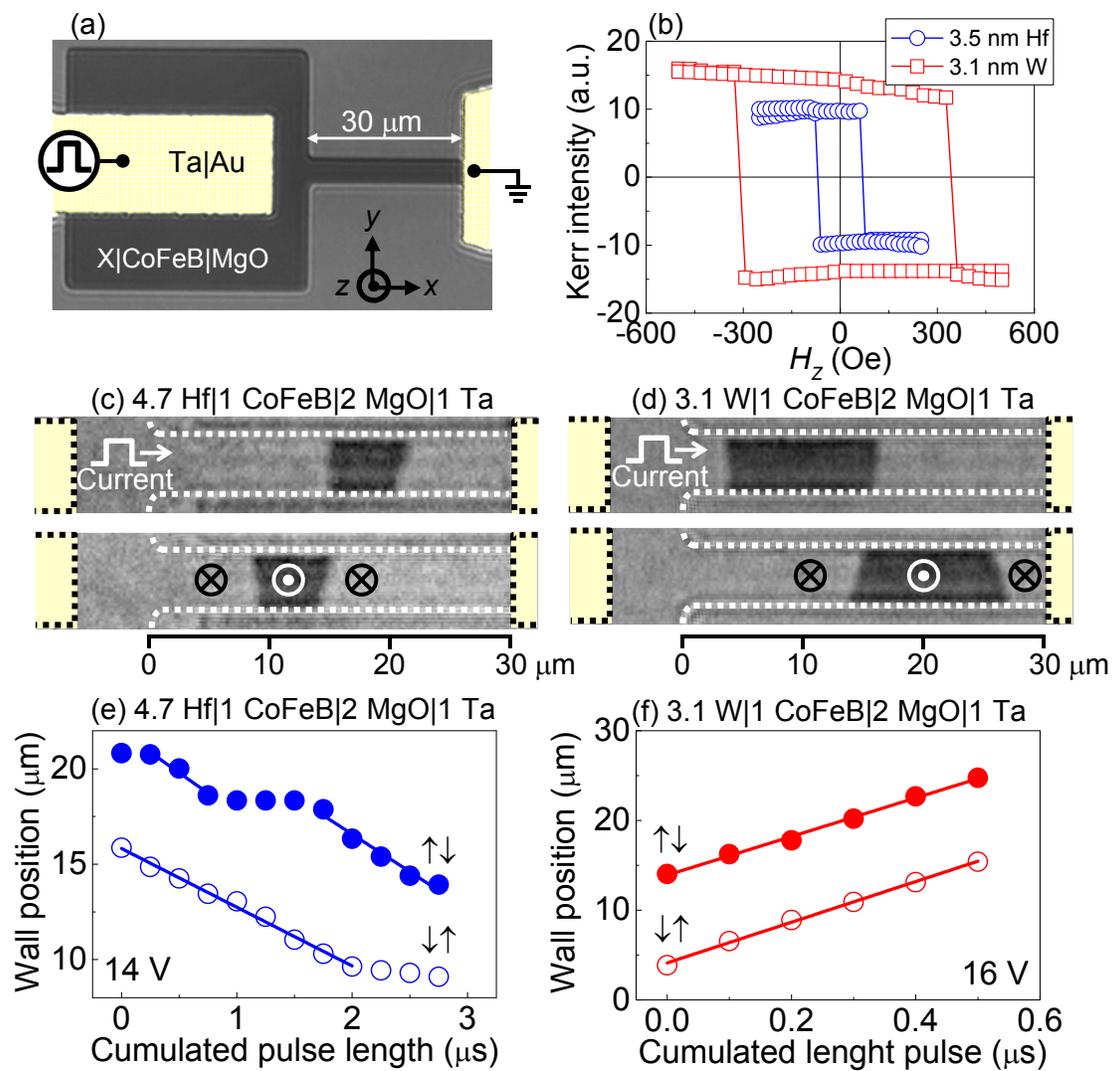

Fig. 2

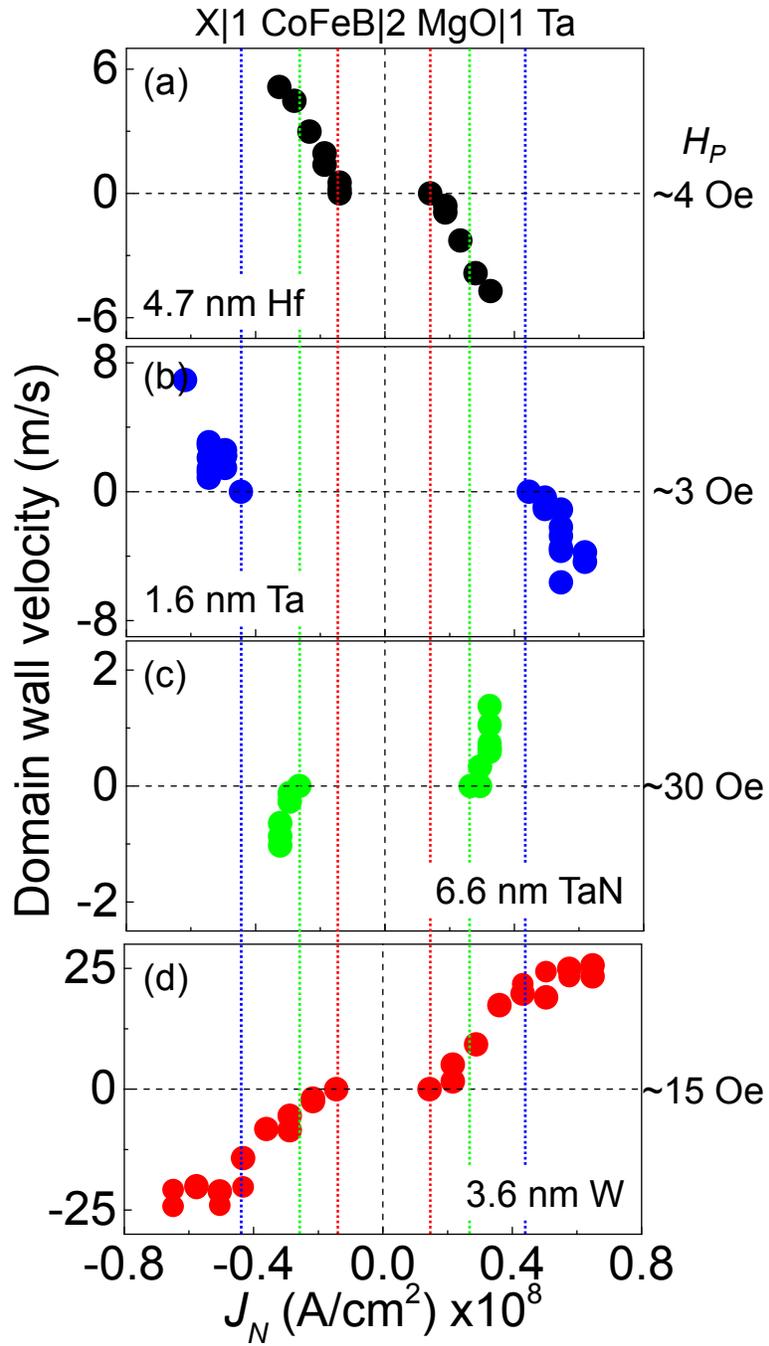

Fig. 3

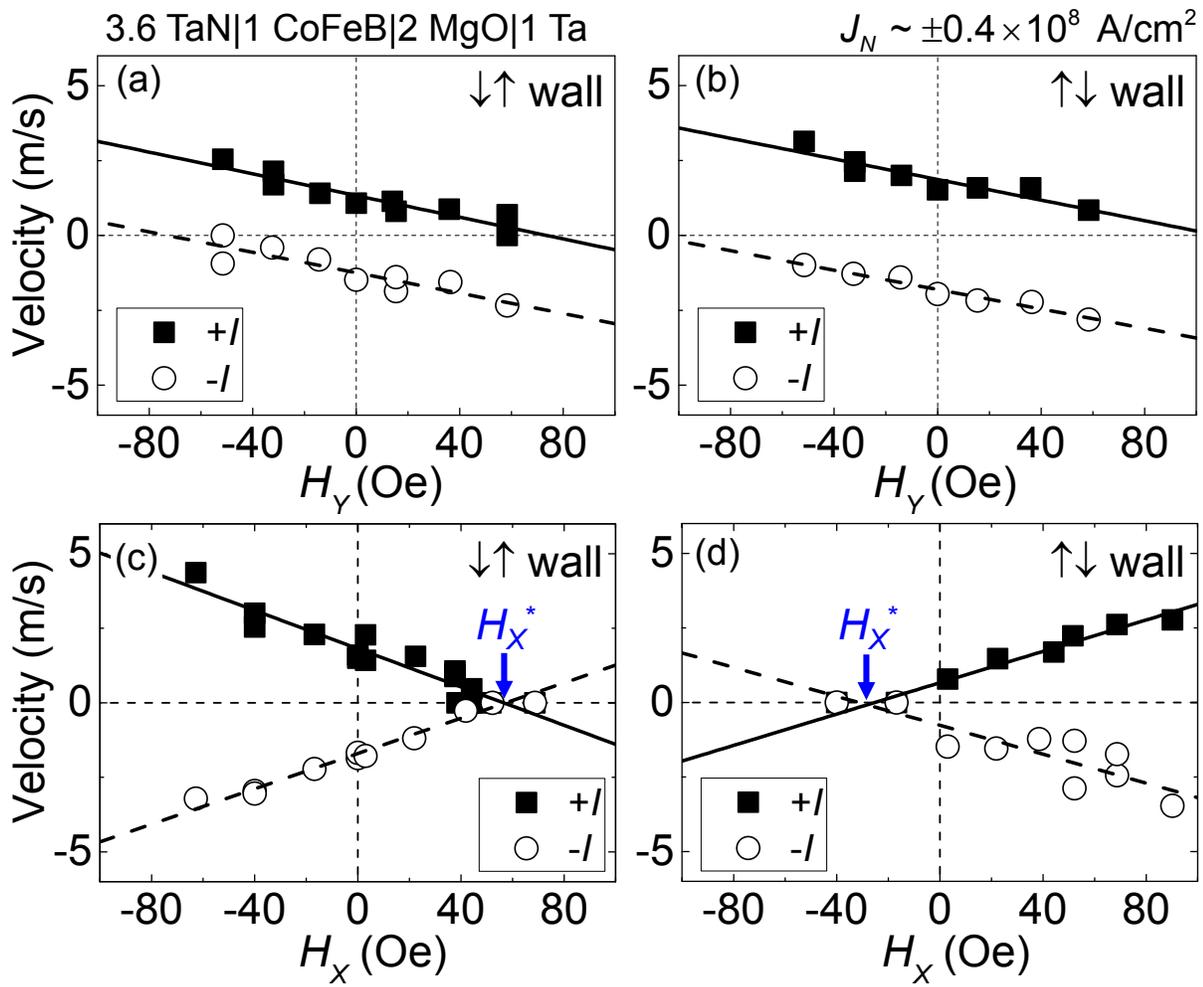

Fig. 4

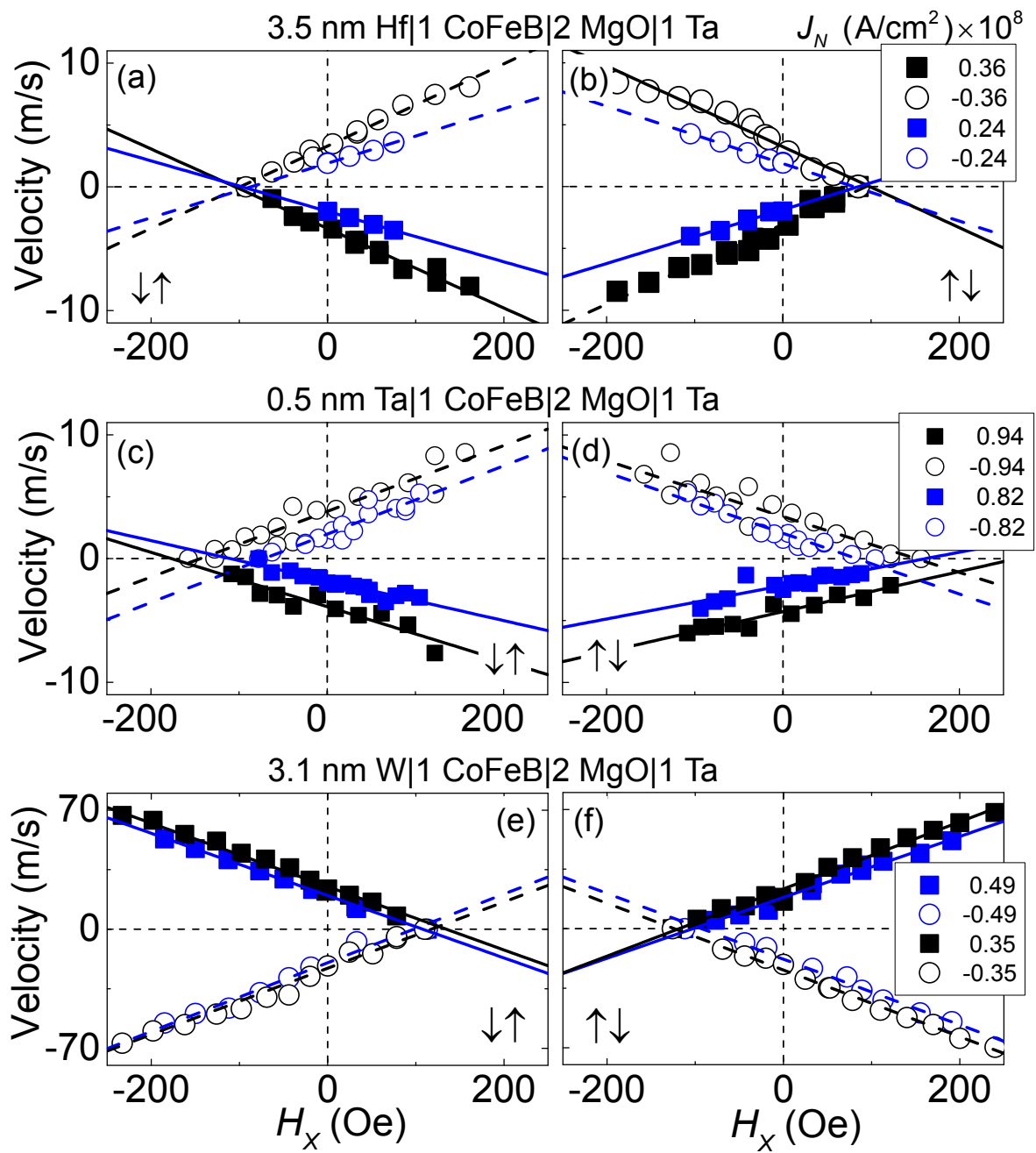

Fig. 5

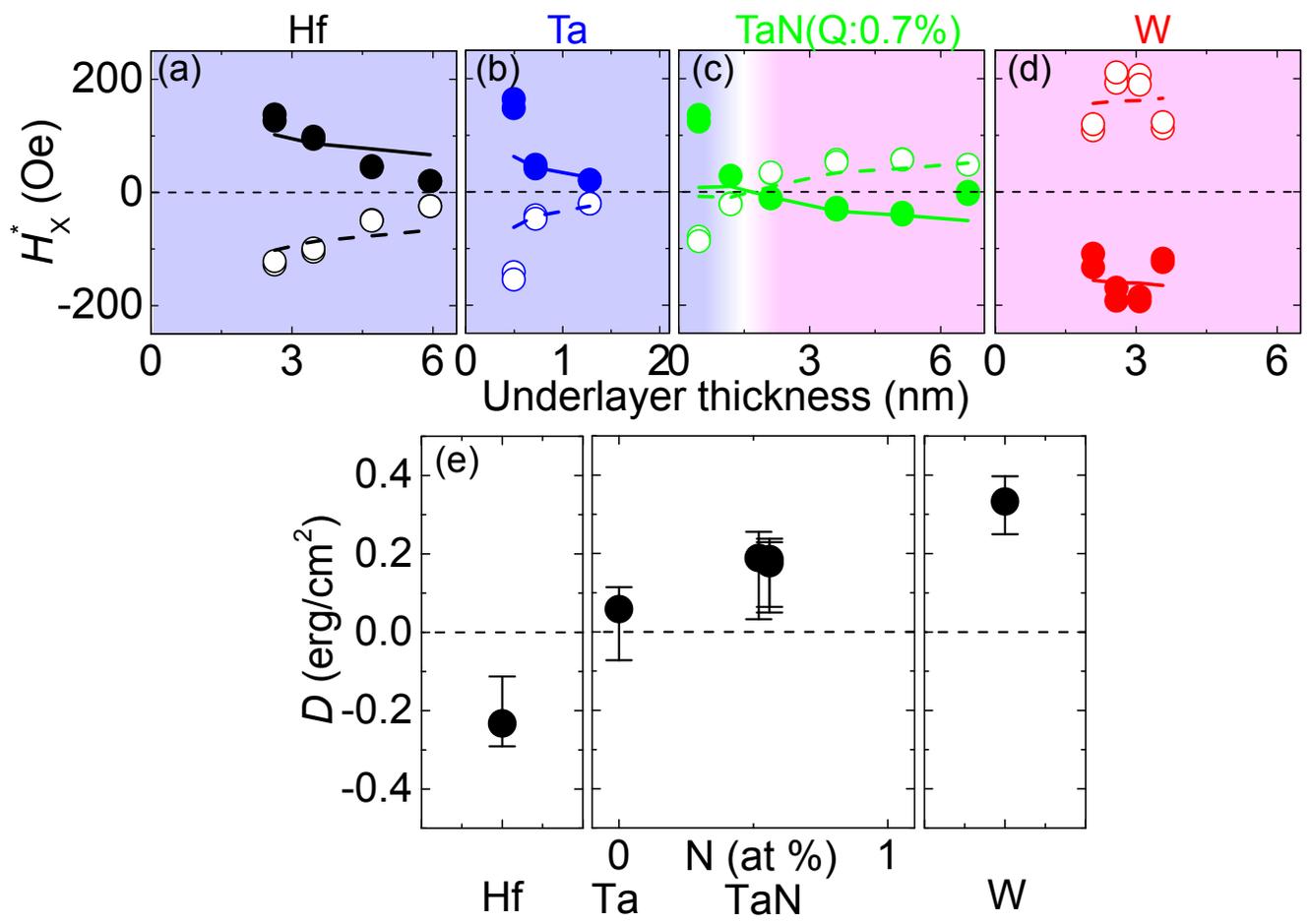

Fig. 6